\documentclass{aa}  
\usepackage{graphicx}
\usepackage{txfonts}
\usepackage[colorlinks=true,linkcolor=blue,urlcolor=blue,citecolor=blue]{hyperref}

\begin{document}

   \title{Resolving the baryon assymmetry with RATS}

   \author{J. Josiek \and 
           M. Bernini-Peron \and
            G. González-Torà \and 
            R. R. Lefever \and
            E. C. Schösser}

   \institute{Technical University of Lurenberg, Département des Heureux, Krankenstrasse 69, 1400 Lurenberg, Listenbourg}

   \date{Submitted February 30, 2025; Accepted April 1, 2025}

  \abstract
   {Current leading theories of physics such as the Big Bang, the standard model of particle physics, and general relativity suggest that the universe should contain an equal amount of matter and antimatter. Yet observations have found a disproportionately large amount of matter, a phenomenon known as the \textit{baryon assymmetry problem}. Since century-old established theories are traditionally impossible to refute, the only possible explanation is that the remaining antimatter is hidden in plain sight and remains to be observed.}
   {We propose the existence of anti-stars to solve the baryon assymetry in our new Reasonable Antimatter Theory of Stars (RATS). In this context, the RATS will create a framework to resolve the traditional tension between observers and theorists, and thus contribute to the peaceful and collaborative spirit of astronomy.}
   {Our method is the firing of neurons in our brains, typically known as a ``thought experiment''. We still have no idea why or how this works, but it must be good because most of science was created this way.}
   {Our results are the result of our methods, which result in some text and the resulting conclusions.}
   {In order to encourage the reader to reach the end of this short paper, we do not want to spoil the conclusions here. Instead, the conclusions will conclude the paper.}

   \maketitle
%

\section{Introduction}

In the beginning, the universe was created\footnote{See Adams (1979)}. Approximately 13.7 billion years later and after an unlikely series of coincidences, matter has assembled to form humans, who now find themselves in the unfortunate position of being the main target audience of this work. 

Matter in the universe has a long and complicated history, but suffice it to say that scientists have understood most of it by now. In this paper, we therefore focus on the much more elusive \textit{antimatter}. Despite the fantastical name reminiscent of any low-budget science-fiction production, antimatter is actually very real, which we know because we have made some and put it in a very expensive box at CERN.\footnote{\url{https://home.cern/science/physics/antimatter/storing-antihydrogen}} Otherwise, it is also flying all around us in small amounts, leaving us undisturbed to focus on more urgent problems.

Antimatter was first proposed by Paul Dirac in 1928, who suggested that every type of fundamental particle possesses an exact opposite analog anti-particle with the same mass and opposite electrical charge. When a particle meets its anti-particle, the pair annihilates and is converted to pure high-energy gamma radiation. The opposite, where a particle-antiparticle pair is produced out of the vaccuum from radiation, is equally possible. Experiments and observations have since confirmed this theory countless times and it is now taken without question.

Another fascinating theoretical interpretation of an anti-particle is that it is a normal particle traveling backwards through time. Understanding this makes use of clever mind-bending tricks that unassuming physics students are routinely subjected to, but that we will not impose on our dear reader. But this way of viewing antimatter is indeed very fun, as demonstrated by its use in entertainment, notably in the film \textit{TENET} (Nolan et al. 2020). Unfortunately, aside from our cinematic canon, scientists have yet to find evidence of large collections of antimatter in the real world. However, the current theory of the Big Bang posits that matter and antimatter should have been created in equal amounts at the beginning of the universe, which sparks our quest for the missing antimatter.

To solve this puzzle, we propose the existence of anti-stars, i.e., stars composed completely of anti-atoms, anti-electrons, et cetera. Since antimatter interacts with the radiation field in the exact same way as normal matter, anti-stars are indistinguishible from stars observationally. Therefore, it is not unreasonable to assume that many stars observed to date are actually anti-stars. This is the core of our new \textit{Reasonable Antimatter Theory of Stars} (RATS). If true, RATS would have the ability to finally bring balance to the universe.

This paper is laid out as follows. In Sect.\,\ref{sec:antisun}, we perform a case study of a prototypical anti-star using our own sun as an example. Sect.\,\ref{sec:ecology} discusses the consequences of a matter planet such as ours orbiting an anti-star, in particular focussing on the effect of the so-called habitable zone. Finally, we place the findings in the wider context of extra-solar stellar populations and discuss the possible observable traces of anti-stars in Sect.\,\ref{sec:extrasolar}, and summarize and conclude our work in Sect.\,\ref{sec:conclusions}.

\section{The anti-sun: a case study}
\label{sec:antisun}

The idea that spawned this paper was originally the question: What if the sun was made of antimatter? As any good scientist knows, the discovery of an interesting and insane question led to a consequent loss of control, and one of us sacrificed something very precious\footnote{A free Saturday.} to obtain an answer. So far, we have not found any scientific literature on this question (except on \textit{reddit.com}), so here is our attempt at filling this serious knowledge gap.

\begin{figure}
    \centering
    \includegraphics[width=0.4\linewidth]{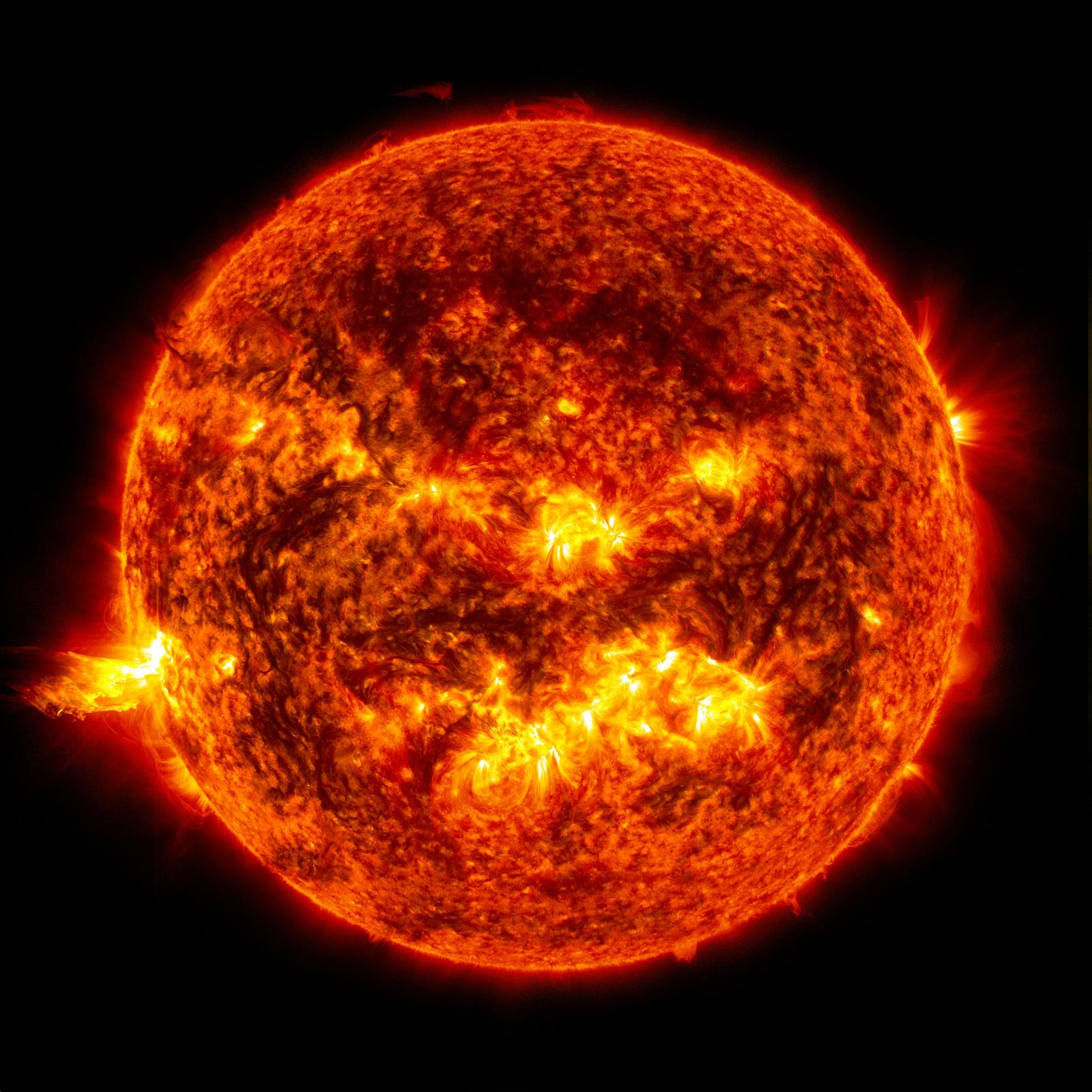}
    \includegraphics[width=0.4\linewidth, trim={105mm 5cm 105mm 5cm}, clip]{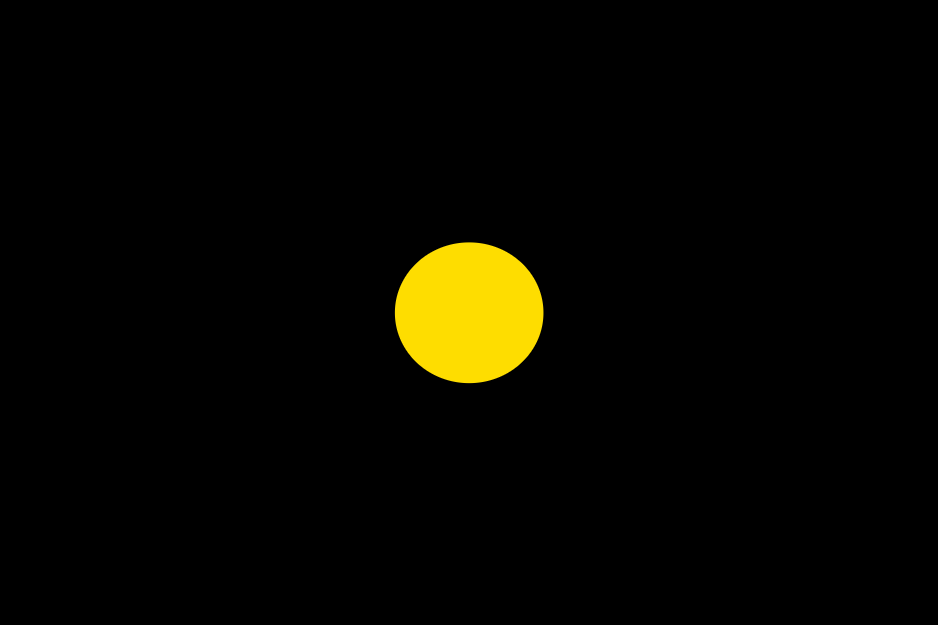}
    \caption{Picture of the sun \textit{(left)} (Credit: NASA), and artist's impression of the anti-sun \textit{(right)} (Credit: Wikimedia)}
    \label{fig:sun}
\end{figure}

To begin, let us consider Fig.\,\ref{fig:sun} depicting both a real picture of the sun as well as an artist's impression of an anti-sun.\footnote{Incidentally, this image was obtained by googling the words ``yellow circle black background'', for which there is actually a jpeg file in the Wikimedia database. We find this a wonderful use of online storage space, and hope this peaceful modern art will get us back into the good graces of future archaeologists as they uncover the rest of today's internet content.} The artist's impression is rather simplistic, since in reality, an anti-sun would look exactly the same as the sun does right now. In fact, we would not notice any immediate changes if the sun was replaced by an anti-sun today. It would still have the same mass (and therefore the same gravity), the same luminosity, the same temperature, and the same tendency to be obscured by clouds only on weekends. As mentioned before, the only distinguishing factor of antimatter is its ability to annihilate matter, and since the sun is in space, there is no matter around it to annihilate. We are therefore safe! Except for one tiny detail.

\subsection{The solar wind}

The solar wind is a flow of ionized particles that are ejected from the sun and dispersed across the solar system. Although there is always a steady outflow of particles from the sun, its intensity varies according to levels of solar activity and can also present as sudden outbursts known as solar flares. The time it takes for the solar wind to reach Earth can be anywhere between 20 minutes to 4 days, which is about the amount of time we have to seek cover.

Considering the scenario where the sun (and thus its wind) consists of antimatter, the anti-particles reaching us will then violently react with the matter in the Earth's atmosphere, producing pure gamma radiation. Most of this energy production will occur in the polar regions due to the Earth's magnetic field deflecting the charged particles in the solar wind. We now perform calculations to determine the magnitude of this effect and assess the ultimate survival of our species in this scenario.

\subsection{Annihilation power from the solar wind}

\begin{figure}
    \centering
    \includegraphics[width=0.8\linewidth]{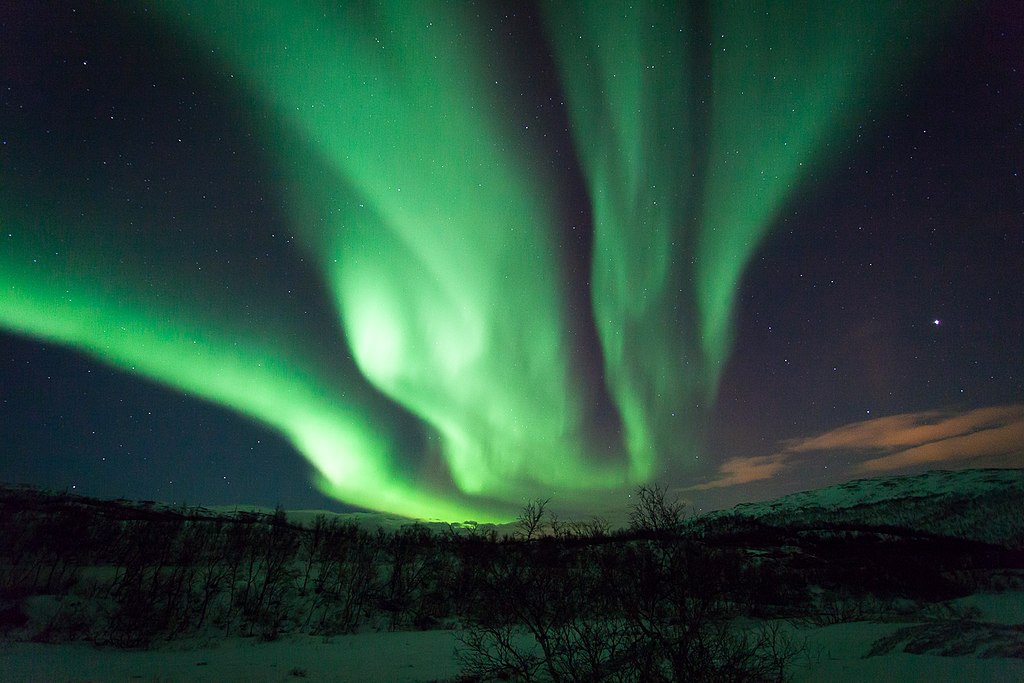}
    \caption{Northern lights produced by the normal solar wind (Credit: Wikimedia)}
    \label{fig:auroraborealis}
\end{figure}

The sun loses about \textit{twenty quadrillionths} of its own mass to the solar wind per year, which is equivalent to one Mount Everest, or in scientific terms,
\begin{equation}
    \dot{M}_\mathrm{total}=2\times10^{-14}\,M_\odot\,\mathrm{yr}^{-1}\text{.}
\end{equation}
Of course, not all of this hits the Earth, so assuming isotropic propagation of the solar wind, our planet receives a fraction of this mass proportional to its cross-sectional surface area. Specifically, if $R$ is the radius of the Earth and $D$ is its distance from the sun, the solar wind received by Earth is
\begin{equation}
    \dot{M}_\mathrm{received} = \frac{R^2}{4D^2}\dot{M}_\mathrm{total} \approx 18\,000\,000\,\mathrm{kg}\,\mathrm{yr}^{-1}\,
\end{equation}
representing an impact of about half a kilogram of solar wind on the Earth every second. Under normal circumstances, this is quite a manageable amount, producing pretty lights (see Fig.\,\ref{fig:auroraborealis}) without any meaningful damage. However, in the anti-sun world, this half a kilogram of anti-solar wind material would violently annihilate with half a kilogram of matter in the Earth's upper atmosphere. The energy produced by this can be easily calculated using the world's most commercialized equation, $E=mc^2$, which yields a total radiative power of
\begin{equation}
    P_\mathrm{antiwind} \approx 2.6\times 10^{16}\,\mathrm{W}\text{.}
\end{equation}

This value is very close to the solar luminosity received by Earth through its normal light ($4\times 10^{16}\,\mathrm{W}$). Essentially this means that turning the sun into antimatter is equivalent to adding a second sun radiating mainly on the poles, and only in gamma rays. 

\subsection{Triggering nuclear decay}

Aside from the purely energetic considerations of the annihilation, we must also consider that the anti-solar wind consists mainly of anti-protons and positrons, but comparatively few anti-neutrons, since the anti-sun is mainly made of the anti-neutron-free anti-hydrogen. As a result, not every atom in the atmosphere will be annihilated equally. Instead, atoms will lose a disproportionate amount of protons and thus become unstable, triggering a chain of nuclear decay.

To estimate the impact of this effect, let us consider a typical atom of nitrogen-14 in the Earth's atmosphere. The nucleus of this atom contains seven protons and seven neutrons and has a binding energy of roughly 100 MeV. The annihilation of the seven protons in this atom with seven anti-protons of the anti-solar wind can be calculated to release about 13.1 GeV of energy, so about 100 times the binding energy of the N-14 nucleus. Thus, even assuming that the entire binding energy can be released or captured upon impact of the wind's anti-particles, we conclude that this effect is negligible compared to the direct annihilation.

Also, let's assume that the overabundance of free neutrons doesn't somehow trigger a run-away chain reaction of nuclear decay in our atmosphere.

\section{Impact on global ecology}
\label{sec:ecology}

Here, we start with the good news. There won't be any sort of global apocalyptic disaster of the sort that can be dramaticized in a disappointing two-hour movie.\footnote{On the other hand, a 7-episode miniseries is possible, but we would like to be consulted by the producers to make sure the science is realistic.} No catastrophic explosion, no immediate boiling or freezing of oceans. The earth will just begin receiving about twice the normal amount of energy from the sun. The anti-sun's wind is also not strong enough to completely evaporate our beloved Earth anytime soon. At a rate of half a kilogram of matter lost every second, it would take more than 100 billion years to annihilate the entire atmosphere, and we simply don't have the patience for that.

Excess gamma radiation will however still have direct impact on human populations. Gamma rays are a type of high-energy ionizing electromagnetic radiation which are extremely toxic to most living species, especially in the doses considered here. Assuming that the radiation is mainly generated above the poles in the upper layers of the thermosphere, located approximately 600 km above the surface, it follows from basic geometry\footnote{``basic geometry'' = we are too lazy to put the math into LaTeX and plot a figure, so instead we gaslight the reader into thinking it is obvious.} that any ecosystem less than $\sim$25 degrees of latitude away from the poles is directly exposed to the gamma rays and therefore uninhabitable. This would necessitate the relocation of about 4 million people currently living above the arctic circle, as well as the few researchers and tourists in Antarctica. 

Let us casually assume that the disruption of the polar ecosystem doesn't really affect us in any way. Then, the gamma rays generated at the poles will simply be absorbed by the upper layers of rock and water on the earth's surface, where it is ultimately converted to thermal energy. This additional dump of energy would certainly disrupt the thermal equilibrium of the surface, and ultimately melt the polar icecaps into the ocean. We don't really know what will happen then, but are currently performing a global experiment to find out. Alternatively, we propose the option for otherwise relocated human beings to just live underneath the rocks that absorb the gamma-rays to protect themselves and use the converted thermal energy as an extra energy source. The additional energy can be used to install a freezer and freeze the melted water back into ice. We are hopeful that future technology will perform this with high efficiency.

Likewise, we are pleased to report that an anti-sun will successfully raise the surface temperature of our cold neighbor Mars, allowing us to simply abandon Earth for a future life of prosperity in our gamma-radiation-proof space suits. We therefore propose to appropriate international funding to convert the sun into antimatter, as we deem this humanity's most realistic chance at successful Martian colonization.

On a future research grant, we will investigate the impact of an anti-sun for the Earth's ecology and to Mars in more detail (Papers II and 3). Paper D will further discuss the economic effects and perspectives of using it as a clean source of energy.

\section{Extra-solar antistars}
\label{sec:extrasolar}

We note that, unfortunately, turning the sun into antimatter simply constitutes an optimistic thought experiment. On the other hand, it is still entirely possible that life would have evolved on Earth anyway if the sun had been an antistar since the beginning. After all, evolution has shown to be incredibly resilient to many environmental difficulties. It is therefore not unreasonable to assume that other planets outside of the solar system are in fact orbiting anti-stars.

Obviously, it could be incredibly difficult to detect anti-stars, since their only signature would be through the interaction with their matter-like environment. Isolated stars with no neighbors and no satellites could well be the universe's hidden reservoir of antimatter. 

Right now, it is unknown how well matter and antimatter are mixed in the universe, so we also cannot predict how frequently to expect an interaction observationally. For example, a galaxy containing both stars and anti-stars would be easier to detect through its gamma-ray excess than pure anti-galaxies. Follow-up studies using the Fermi Gamma-Ray Space Telescope on targets across various cosmic scales are required in order to shed more light on this question. Unfortunately, our previous requests for observing time have been rejected so far, and thus our theory will become legend, and legend will become myth.

\section{Conclusion}
\label{sec:conclusions}

Like discussed previously, the main conclusion of this work is that the sun is probably not made of anti-matter. This was perhaps obvious, but we have not found this explicitly stated in the scientific literature and would therefore appreciate a citation should this fact be referred to again in future work.

Definitive questions still remain about the nature of RATS, which we introduced in the introduction and then conveniently ignored until now. RATS is an acronym which stands for \textit{Reasonable Antimatter Theory of Stars}, which we have established throughout this work. It is also the word \textit{star} written backwards, alluding to the opposite-but-equal nature of antimatter. It also allows us to write the sentence, ``the solution is RATS'' in all seriousness in a paper. For these reasons, we believe it to be a suitable acronym for use in professional astronomy. Further studies are needed to determine if it is also a suitable theory.

\begin{acknowledgements}
      This work was produced without the knowledge or approval of our employer, the German Federal State of Baden-Württemberg, but was indirectly funded through them anyway, by way of our salary which we mostly spend on our ambitious weekend plans. We wish to acknowledge the anonymous self-proclaimed ``experts'' whose daring confidence taught us that neither logical methodology nor reliable sources are requirements to publish a creative theory. We gratefully acknowledge our own remaining ability to formulate coherent sentences without the use of generative language models, and certify that no machine (other than the coffee machine) was harmed in the production of this text. \\ \\ Finally, and in all seriousness, we could not have written this paper without the many lively and light-hearted discussions that happen in our group on an almost daily basis. It is a tribute to the value of humor, friendship and fellowship in and around the science community.
\end{acknowledgements}

\end{document}